# Differences in personal and professional tweets of scholars

Timothy D. Bowman

timothy.bowman@umontreal.ca
École de bibliothéconomie et des sciences de l'information (EBSI), Université de Montréal, Montréal, QC (Canada)

**Abstract**

**Purpose** – The purpose of this paper is to show that there were differences in the use of Twitter by professors at AAU schools. Affordance use differed between the personal and professional tweets of professors as categorized by turkers. Framing behaviors were described that could impact the interpretation of tweets by audience members.

**Design/methodology/approach** – A three phase research design was used that included surveys of professors, categorization of tweets by workers in Amazon's Mechanical Turk, and categorization of tweets by active professors on Twitter.

**Findings** – There were significant differences found between professors that reported having a Twitter account, significant differences found between types of Twitter accounts (personal, professional, or both), and significant differences in the affordances used in personal and professional tweets. Framing behaviors were described that may assist altmetric researchers in distinguishing between personal and professional tweets.

**Research limitations/implications** – The study is limited by the sample population, survey instrument, low survey response rate, and low Cohen's κ. Practical implications – An overview of various affordances found in Twitter is provided and a novel use of Amazon's Mechanical Turk for the categorization of tweets is described that can be applied to future altmetric studies.

**Originality/value** – This work utilizes a socio-technical framework integrating social and psychological theories to interpret results from the tweeting behavior of professors and the interpretation of tweets by workers in Amazon's Mechanical Turk.

**Paper type** – Research paper

# 1 Introduction

The current conversations in both popular media and academic discourse surrounding the use of the social web places pressure on various groups within academia –such as scholars, universities, and funding agencies–to consider a variety of behaviors, policies, and regulations related to the production, consumption, disclosure, and dissemination of information within these environments. Interaction in computer-mediated environments is a ubiquitous aspect of day-to-day life (Mitzlaff et al., 2013) and as such researchers look to examine the digital traces (Lazer et al., 2009) left behind in social media by participants (e.g. scholars), in an attempt to analyze phenomena such as utilizing and maintaining social capital (Hoffmann et al., 2014), managing impressions (Haustein et al., 2013; Veletsianos, 2012), influencing science (Priem, 2010; Priem et al., 2010, 2012), and consuming and disseminating scholarly information (Bowman et al., 2013; Schroeder et al., 2011). In addition, universities and organizations look to this research to assist them with evaluating scholarly production (Bar-Ilan et al., 2012) and crafting social media use policies (Duque and Pérez, 2013; Hank et al., 2014; Lough and Samek, 2014) as the boundaries between public and private continue to blur.

In the case of social media this blurring of boundaries is particularly important as the information within these environments can be archived, searched, reproduced, and viewed by vast invisible audiences (boyd, 2011). Twitter is an important context for researchers studying social media metrics (or so-called altmetrics) because analyses have found that scientific articles are frequently shared in this environment (Haustein et al., 2014a; Holmberg and Thelwall, 2014) and, opposed to reference managers like Mendeley, include potential audiences outside academia. In general, social media metrics researchers are trying to identify and distinguish between the dissemination of, and engagement with, scientific content on social media, in order to determine whom, how, and why users share and consume academic discourse in these environments, to compare the activity with traditional bibliometric measures, and to determine what type of impact this dissemination and engagement in social media has on both academia and the general public. As Holmberg and Thelwall (2014, para. 1) argue, "it seems that social media are triggering another evolution of scholarly communication".

With the scholarly use of these social media platforms comes an increase in the potential for audience members to interpret messages in unexpected ways. For example, scholars have been placed on leave, disciplined, or had their job offer rescinded (Berrett, 2010; Herman, 2014; Ingeno, 2013; Jaschik, 2014; Rothschild and Unglesbee, 2013) for messages posted within the context of social media (See Figure 1). Because of issues like these, universities are involved in a continuous cycle of crafting rules and norms intended to distinguish between appropriate and inappropriate social media use (Duque and Pérez, 2013; Hank et al., 2014; Lough and Samek, 2014; Sugimoto et al., 2015). It is critical that we examine how it is that scholars are communicating and sharing information within these contexts (Haythornthwaite and Wellman, 1998), how their interactions are affected by the affordances (Gibson, 1977) available in the social media platform, and how their messages are interpreted by the vast potential audience members to whom the information is being disseminated.

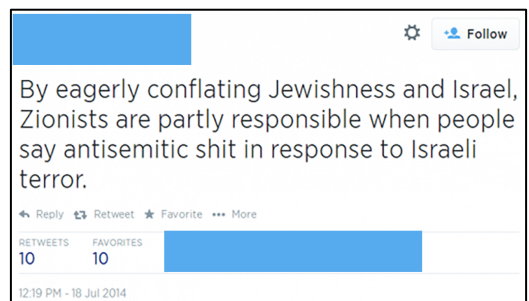

**Figure 1** Scholar's controversial tweet made on July 18, 2015.

This research examines the ways in which scholars can utilize affordances (i.e. a quality of an object allowing for some action in a context) and frame interactions (i.e. add meaning to an act in order to aid the understanding of the act in its context) in order to manage communications from the perspective of their personal and professional selves as they communicate on Twitter. This socio-technical framework combines Goffman's (1974) frame analysis model, Goffman's (1959) impression management model, and Gibson's (1977) conception of affordance. Using this framework, the following research questions are addressed:

> *RQ1*. How does Twitter use by university professors differ by department, gender, age, ethnicity, and academic age?
>
> *RQ2*. How does the tweeting behavior and affordance (#hashtags, @user mentions, URLs, and RT:retweets) use of university professors differ across personal and profession tweets by department, gender, age, academic age and Twitter activity?
>
> *RQ3*. How can university professors frame tweets using affordances so that their audience is able to distinguish between personal and professional tweets?

The primary motivation for this work is to examine how affordances may be used to reinforce frames applied to tweets by an audience as scholars manage their personal and professional impressions on Twitter. A secondary motivation is to examine how often scholarly tweets are misinterpreted by audience members as personal when they are professional or vice versa. A tertiary motivation is to introduce the use of Amazon Mechanical Turk (AMT) as a viable methodological possibility for the future evaluation of social media data.

The results presented here represent a proof of concept for future analyses of scholarly participation on social media; examining affordance use and framing behaviors using this framework can assist universities and organizations to distinguish between the personal and professional communications of scholars; allow altmetrics researchers to answer questions related to who, how, and why users are sharing and discussing scholarly contents on Twitter; and assist academics themselves with framing their own tweets. The study applies a mixed methods approach by combining surveys and categorization techniques to collect, label, and analyze the tweets and affordances used by scholars within personal and professional tweets. The contribution of this work stems from the use of social theory to interpret scholarly communication in social media, the use of AMT to categorize tweets, and provides further insight into how scholars are using Twitter.

## 2 Theoretical Framework

Beginning in the late 1990s, scholars such as Baldwin (1998) contemplated that academia was being affected in various ways by emerging computer-mediated environments, arguing that they allowed for, among other things, insight into the way academic life was evolving. In addition to scholars themselves, organizations both within and outside of higher education presumably also want to know what types of effects social media platforms are having on scholarly communication, dissemination, and information exchange. Davis III, et al. (2012) warned that it is critical for researchers to investigate how organizations, such as organizations in higher education, are incorporating, monitoring, and normalizing the use of social media by their employees. There have also been studies (Haustein et al., 2013; Veletsianos, 2012) that point to issues facing scholars as they manage their impressions within social media environments. From this and other work it is clear that the use of social media by scholars is having an impact on their once invisible backstage activity, as Priem (2014, p. 264) argues, by bringing "the background of scholarship […] out onto the [front] stage."

This notion of backstage activity is what Goffman (1959) formalized in his sociological framework describing and investigating face-to-face interaction. He used dramaturgical concepts such as stage and actor to describe the ways in which people interact on a daily basis. Goffman's insights into social interaction at the individual level have been used by scholars across multiple disciplines (Leary and Kowalski, 1990) to discuss and explore how persons interact with one another, while taking into account the context in which the interaction occurs and the symbols and props used to help the audience understand the roles being enacted by the actors. In a later work, Goffman (1974) expanded upon this dramaturgical model by discussing the frames that are used by actors to make sense of interactions; Goffman considered a frame to be a (potentially multi-) layered set of norms, rules, props, and experiences used by individuals to interpret and comprehend what is occurring during any interaction. While Goffman focused solely on the physical constraints used to interpret communication within an environment, other scholars (Meyrowitz, 1990; Miller, 1995) have intuitively extended Goffman's work to examine interaction in the digital environment by focusing on information constraints instead of physical constraints.

Gibson (1977) defined an affordance as a functional attribute of an object within a specific context (or niche). Affordances are useful to describe the attributes (instead of physical characteristics such as weight, color, height, etc.) of digital spaces and have been used by others (Boyd and Ellison, 2008; Gilpin, 2011; Murthy, 2013; Papacharissi, 2011) to describe interaction within social media environments. Combining Goffman's (1959, 1974) ideas and Gibson's (1977) concept provides for unique insight into the ways in which scholars make use of the affordances in social media environments like Twitter to differentiate their personal and professional tweets for their audience. It aids in understanding questions proposed in the area of social media metrics and scholarly communication regarding who and why Twitter users are consuming and disseminating scientific information in their tweets.

# 3 Methods

This exploratory research was carried out in three distinct phases from January 2014 through October 2014. Each phase is described below.

*3.1 Phase One: Survey*
A 19-question survey was designed to gather information about Twitter use, affordance use, and basic demographic information. The survey was pilot tested with one faculty member and two graduate students from the field of Information and Library Science and built using Qualtrics survey software. In consecutive pilot tests it took participants approximately six minutes to answer all of the 19 questions if the user had one Twitter account, and up to 10 minutes when a user had five accounts. A sampling frame of 16,862 professors was created by gathering the available demographic information of full-time assistant, associate, and full professors from eight disciplines (Physics, Biology, Chemistry, Computer Science, Philosophy, English, Sociology, and Anthropology) at 62 Association of American Universities (AAU) member institutions found on publically available departmental web sites between September 2013 and January 2014. Adjunct professors, doctoral students, and other members of the departments who were listed as something other than full-time faculty were not included. The sample is a purposive sample in that only members of AAU institutions from select departments are included, a cluster sample in that institutional web sites were used to gather groups (i.e. individual professors) for the unit of analysis, and a convenience sample in that only the information of persons found on these departmental web sites were gathered.

The survey was built and delivered using Qualtrics survey software. Because of limitations on the number of surveys delivered via e-mail in Qualtrics, the survey was sent to the sample of scholars in two batches: the first batch was sent on January 26, 2014 to 9,677 scholars from 39 universities and the second batch was sent on February 3, 2014 to 7,185 scholars from the remaining 23 universities. A reminder e-mail was sent to the first group on February 11, 2014 and to the second group on February 16, 2014. The combined survey invitations totaled 16,665 and the total number of surveys started was 1,960 for a response rate of 8.5 percent.

*3.2 Phase Two: Amazon Mechanical Turk Tweet Categorization*
Of the 1,910 respondents who answered at least one question in the survey, a total of 613 respondents answered "Yes" to having at least one Twitter account. Of this group, valid Twitter account handles were verified for 391 scholars resulting in 445 unique Twitter accounts; verification occurred by searching Twitter and Google for the scholar's name, e-mail address, university, university location, and/or Twitter handle(s) reported in the survey. The final 445 Twitter accounts used in this work include both individual accounts and lab accounts. Tweets from lab accounts were included in this study because the messages are considered to be sent on the scholars' behalf – a similar rationale was used by Hemphill et al. (2013) when examining political tweets.

The Twitter profile information and a sample of tweets from each of the 445 accounts were collected through the Twitter API using a PHP script on May 19, 2014. The Twitter API enforces a retrieval limit of 3,200 tweets per user, thus the final total collected from the 445 Twitter accounts was 289,934 tweets from a possible 585,879 sent. A long tail distribution was found when examining the number of tweets by account, as there were many accounts that had a far fewer mean tweets per day average (TPD Average = total tweets/Days account open) than the mean of 0.88 TPD (median of 0.16 TPD, no mode).

Because of this variation in the data, a stratified proportionate sampling technique was utilized to obtain 75,000 random tweets for use in AMT. To obtain the 75,000 random tweets for use in AMT, the scholars were first divided into ten groups ranging from infrequent to intense Twitter use as determined by TPD averages per scholar (as seen in Table I). A random sample of tweets was then gathered using the percentages of total tweets per group (represented in the 'Percentage of Total Tweets' column in Table I) as the subset sample size, which was then multiplied by 75,000 to obtain a final sample of tweets for each group. For example, group ten (i.e. intense users) accounted for 10.02 percent of the 289,934 total tweets collected, therefore 10.02 percent of 75,000 was randomly chosen from group ten's tweets, which resulted in 7,518 tweets collected from this group.

**Table 1** Grouping of Scholars by Tweets per Day Average. *BOLD type indicates those scholars included in phase three.

| Group Name | Average Tweets per Day (TPD) | Total Scholars in Group | Total Tweets Collected | Percentage of Total Tweets | Tweets Used in AMT |
|---|---|---|---|---|---|
| **TEN (intense)** | **8 to 24** | **9** | **29,064** | **10.02%** | **7,518** |
| **NINE** | **5 to 8** | **8** | **25,863** | **8.92%** | **6,690** |
| **EIGHT** | **4 to 5** | **6** | **19,321** | **6.66%** | **4,998** |
| **SEVEN** | **3 to 4** | **10** | **24,532** | **8.46%** | **6,346** |
| **SIX** | **2.5 to 3** | **10** | **25,508** | **8.80%** | **6,598** |
| **FIVE** | **2 to 2.5** | **10** | **22,195** | **7.66%** | **5,741** |
| **FOUR** | **1.5 to 2** | **13** | **23,018** | **7.94%** | **5,954** |
| **THREE** | **1 to 1.5** | **29** | **43,831** | **15.12%** | **11,338** |
| TWO | 0.5 to 1 | 33 | 30,463 | 10.51% | 7,880 |
| ONE (infrequent) | < 0.5 | 317 | 46,139 | 15.91% | 11,935 |
|  |  | *445* | *289,934* | *100.00%* | *75,000* |

The final sample of 75,000 tweets was placed in the AMT application in order to determine how a random audience would categorize them. AMT is a marketplace that can be used to crowdsource tasks relating to a variety of concepts including marketing, opinion, and research; a person (or company) known as a requestor can create tasks, called Human Intelligence Tasks (HITs), and pay turkers (Amazon calls them providers, but they are commonly known as turkers) to perform a set number of these tasks.

A total of 12,056 HITs were created that presented up to seven tweets to each turker for categorization using a standard HTML template within AMT. To ensure data integrity, a control question was assigned to each HIT specifying that the turker had to categorize a sample tweet into a specific category or their results would not be valid and they would not be paid. With each HIT completed the turker earned a wage of $0.10 for categorizing up to seven tweets as either " Personal" , " Professional" , " Unknown" , or " Non-English" . The categories of " Unknown" and " Non-English" were included because a preliminary analysis of the sample by the author found that several tweets were written in various languages and that there were tweets that contained nothing but punctuation or other miscellaneous characters where there was no clear way to categorize them. Three distinct turkers were assigned to each HIT; the turkers were self-assigned and were required to meet the criteria specified in the description to complete a HIT. To be qualified for this work, a turker was required to have previously completed at least 10,000 HITs and have an average HIT approval rate of 99 percent. The turkers took an average of 1 minute and 28 second to complete each HIT.

The turkers fully agreed (three out of three) on the categorization of 34,969 tweets (47 percent) across four categories: personal (n. 27,264), Professional (n. 6,810), non-English (n. 766), and unknown (n. 129). Turkers partially agreed (2 out of 3) on the categorization of 37,355 tweets (49 percent) across the four categories: personal (n. 19,403), professional (n. 15,692), non-English (n. 262), and unknown (n. 1,993). Finally, turkers disagreed (0 out of 3) on 2,674 tweets (4 percent). The AMT tweet categorization results demonstrated that at least two out of three turkers agreed on the categorization of 96 percent of all tweets categorized. The control question was answered correctly in all of the agreed and partially agreed HITs.

*3.3 Phase Three: Follow-up Survey and Tweet Categorization*
The follow-up survey contained a maximum of six questions and was designed to gather information about affordance use across personal and professional tweets, affordance use in profile information, and to have scholars categorize five of their own publically available tweets as either "Personal" or "Professional". The tweets presented to the respondents were chosen randomly from phase two where turkers fully agreed (3 out of 3) on the categorization of the tweet (except for two cases in which only two professional tweets were available; in this case tweets were randomly chosen from the partial agreement group of tweets for that specific scholar). Three professional and two personal tweets were randomly selected and then randomly ordered and presented to the scholars in the same format as they were shown to the turkers.

A sample of 95 scholars from 25 of the 62 AAU universities who, on average, tweeted at least once per day (from group three through ten, Table I) were included in this phase of work because these scholars demonstrated consistent tweet activity and therefore have a better understanding of the affordances available in Twitter. These respondents were invited to participate on Wednesday, October 22, 2014 and a reminder e-mail was sent on October 27, 2014. There were 66 respondents who started the survey and 57 who completed for a response rate of approximately 63 percent.

## 4. Results

*4.1. Phase One: Survey*

*4.1.1. 1,910 Survey Respondents.*

Those professors reporting having a Twitter account (n=613; 32 percent) were compared against those without an account by department, academic age, ethnicity, academic title, and gender. Results show that there are statistically significant differences regarding Twitter uptake by ethnicity, academic age, department, and title. Professors from computer science (50 percent) reported the highest proportion of Twitter accounts, whereas professors from chemistry (21 percent) reported the lowest (Figure 2). Performing a χ2 test by department showed a strong relationship between department and having a Twitter account, χ2(7, n=1,910)=0.182, p=0.0005, Cramér's V=0.182. Respondents were asked how long they had been a faculty member at a university; the data indicate that there was a strong relationship between academic age–reported as six years or less, seven to nine years, and ten years or more–and having a Twitter account. Professors reporting having been a faculty member seven to nine years had the highest percentage of accounts (41 percent) followed closely by those reporting as six years or less (39 percent), whereas only 25 percent of those having been a faculty member ten years or more reported having a Twitter account. A χ2 test examining academic age found a strong relationship between academic age and having a Twitter account, χ2(2, n=1,910)=0.217, p=0.0005, Cramér's V=0.217.

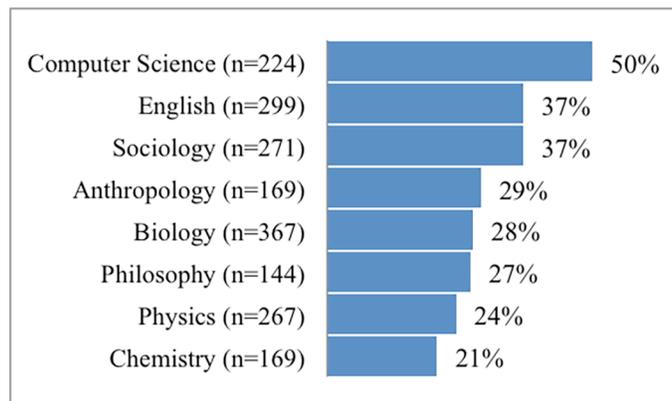

**Figure 2** Percentage of Twitter Account Holders by Department

Grouping ethnicity (n=1,910) by white/Caucasian and non-white found that 24 percent of white/Caucasians reported having a Twitter account and only 8 percent of non-whites had an account. There was a strong relationship between ethnicity and having a Twitter account as a χ2 test revealed, χ2(1, n=1,910)=−0.140, p=0.0005, Cramér's V=0.140. Scholars who were full professors made up 42 percent of the scholars with Twitter accounts compared with 29 percent of both assistant and associate professors. A strong relationship was found between academic title and having a Twitter account, χ2(2, n=1,910)=0.154, p=0.0005, Cramér's V=0.154. Lastly, results by gender (n=1,824) found that 28 percent of males reported having an account compared with 33 percent of females, a χ2 test revealed that while significant, the result was a neglible relationship—χ2(2, n=1,824)=0.066, p=0.018, Cramer's V=0.18.

Respondents (n=1,639) were also asked what social media tools they used besides Twitter. The results indicated some variety amongst the scholars as Facebook (70 percent), LinkedIn (58 percent), and Google+ (50 percent) were by far the most reported general social media platforms used by the scholars. Academic social media tools were also given as options and the results indicate that social networks ResearchGate (26 percent) and

Academia.edu (22 percent) had a much higher use reported than reference manager Mendeley (7 percent).

*4.1.2 613 Scholars with Twitter Accounts.*
Most scholars indicated that they used their Twitter account both personally and professionally (42 percent; n=230), while some indicated they used it for distinctly personal (29 percent; n=164) and others for distinctly professional tweeting (29 percent; n=159). These results reiterate the point that many scholars are communicating both personally and professionally using the same Twitter account.

A χ2 test of significance was run for associations between ethnicity, age, academic age, department, academic title, and gender and type of Twitter use; it was found that personal, professional and mixed use of Twitter did not differ by ethnicity, academic age, gender, and academic title. With regards to department (Figure 3), philosophers (44 percent) indicated a higher number of personal-only accounts, whereas English professors (60 percent) reported the highest number of both personal and professional accounts. Scholars from sociology and computer science reported the highest number of professional-only accounts (34 percent). Differences were significant between departments, χ2(14, n=553)=0.279, p=0.0005, Cramer's V=0.197.

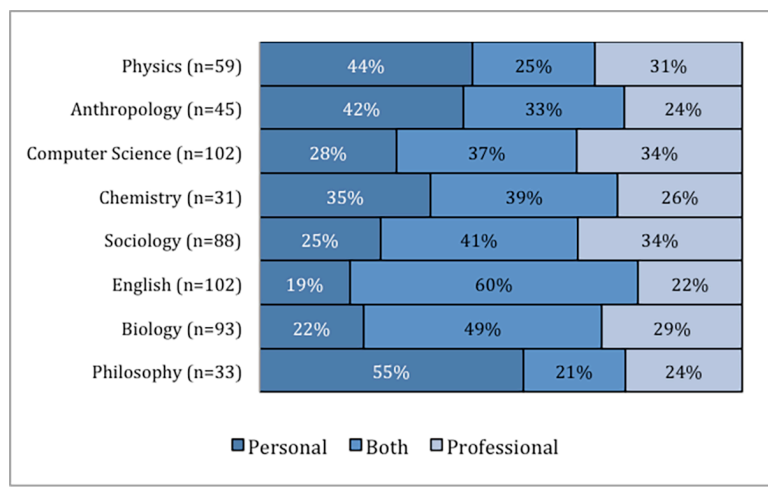

**Figure 3** Differences by department looking at the type of Twitter account chosen by professors as either personal, professional, or both.

There were also differences between age groups–reported as 35 and under, 36 to 45, 46 to 60, and 61 and over–and type of Twitter use. Those in the 35 and under age group indicated they considered their accounts only professional more than expected, those in the 36 to 45 group chose both personal and professional more than expected, both the 46 to 60 and 61 and over groups chose personal only more than expected. The results from a χ2 test found that there was a moderate association between age and Twitter use (χ2(6, n=508)=0.197, p=0.003, Cramer's V=0.139). The data illustrate that as scholars become older they tend to use their Twitter accounts less for both personal and professional communications, with a higher percentage using their account simply for personal communications. Scholars between 36 and 45 years old reported the most combined (personal and professional) accounts, while those under 35 indicated they use their accounts in a professional-only manner.

*4.1.3 Twitter Data From 391 Scholars.*
As mentioned in the methods section, the most recent 3,200 tweets and profile information for each account were collected for the 395 professors whose accounts were verified. When looking at median differences between departments, it was found that English professors had a higher median of followers (294), friends (150), and total tweets (410) than all other departments. Professors from chemistry had the lowest median number of followers (43), professors from physics had the lowest median number of friends (33), and professors from philosophy had the lowest median number of total tweets (39).

The four primary affordances found in the tweets were compared across departments (shown in Table II) and the results indicated clear differences in the way scholars made use of affordances. Sociologists were found to have the most occurrences of hashtags (7.4 percent) and user mentions (20 percent), whereas URLs (1.7 percent) were found to occur most in tweets by philosophers and retweets (291) occurred most amongst tweets by English professors. A one-way ANOVA was run comparing the differences across departments in hashtags, mentions, URLs, and retweets, but the assumption of homogeneity of variances was violated for user mentions (p=0.009), and URLs (p=0.004), as assessed by Levene's test for equality of variances. Because of these violations, a one-way Welch ANOVA was run. It was found that the amount of URLs used (Welch's $F(7, 4.131)=125.019$, p=0.0005) was statistically significantly different for the eight different departments. Differences between departments regarding hashtag use, user mentions, and retweets were not found to be significant.

**Table 2** Mean average of affordance use by department.
A=Anthropology, B=Biology, C=Chemistry, CS=Computer Science, E=English, PH=Philosophy, PY=Physics, S=Sociology

|          | A     | B     | C     | CS    | E     | PH    | PY   | S    | Average |
|----------|-------|-------|-------|-------|-------|-------|------|------|---------|
| Hashtags | 4.4%  | 5.5%  | 5.2%  | 5.2%  | 4.9%  | 4.6%  | 6.4% | 7.4% | **5.5%** |
| URLs     | 0.7%  | 1.2%  | 0.3%  | 1.1%  | 0.5%  | 1.7%  | 0.8% | 1.1% | **0.9%** |
| Mentions | 11.6% | 16.3% | 12.9% | 9.2%  | 13.4% | 10.6% | 13%  | 20%  | **13.4%** |
| Retweets | 241   | 273   | 137   | 244   | 291   | 171   | 124  | 205  | **211** |

The average of the mean tweets-per-day (TPD) per scholar was calculated and compared for all independent variables. Although scholars from philosophy (1.96) had the highest average of mean TPD as compared to chemists (0.52) and physicists (0.52), who demonstrated the lowest average of mean TPD, difference of tweeting activity was not statistically significant (Welch's $F(7, 1.537)=115.843$, p=0.162). Overall the scholars from the social sciences (1.40) averaged a higher mean TPD than the scholars from the natural sciences (0.61). The average tweeting activity did not differ by gender.

*4.2 Phase Two: Amazon's Mechanical Turk*
A closer inspection was made on the tweet content from the sets of personal and professional tweets in which the Turkers fully agreed (n=34,074). Mann-Whitney U tests were run to determine if there were differences in user mention use, hashtag use, and URL use between personal and professional tweets. Distributions of the affordances for personal and professional tweets were similar, as assessed by visual inspection. Median user mention use was statistically significantly different between personal (Mdn=2,160) and professional tweets (Mdn=1,469), U=78,129,900.5, z=-20.252, p=0.0005. Regarding the use of hashtags, it was found that median hashtag use was statistically significantly different between personal (Mdn=363) and professional tweets (Md=436), U=99,903,773, z=9.737, p=0.0005. Finally, median URL use was statistically significantly different between personal (Mdn=612) and professional tweets (Mdn=704), U=102,801,773, z=13.729, p=0.0005.

*4.2.1 Affordance Use in Agreed Personal Tweets.*
The mean average of affordance use was compared across departments for the agreed set of personal tweets (three of three turkers agreed on categorization). Chemists (0.24) were found to have the highest mean of tweets with at least one hashtag, while philosophers (0.09) had the lowest mean.

Sociologists (0.22) had the highest mean of tweets with at least one URL, while chemists (0.11) and physicists (0.11) had the lowest mean. An analysis of the quantities of user mentions found that biologists (0.76) had the highest mean of tweets with at least one user mention, while chemists (0.62) had the lowest mean. Finally, biologists (0.26) had the highest mean of tweets that were retweets and physicists (0.12) had the lowest. A $\chi^2$ test for association was conducted between department and affordance use. All expected cell frequencies were greater than five. There was a statistically significant association between department and hashtag use, $\chi^2(7\ n=27,264)=0.125$, p=0.0005, Cramer's V=0.125, between department and URL use, $\chi^2(7\ n=27,264)=0.093$, p=0.0005, Cramer's V=0.093, between department and user mention use, $\chi^2(7\ n=27,264)=0.096$, p=0.0005, Cramer's V=0.096, and between department and retweeting, $\chi^2(7\ n=27,264)=0.095$, p=0.0005, Cramer's V=0.095.

*4.2.2 Affordance Use in Agreed Professional Tweets.*
A similar analysis (as to 4.2.1) was performed on the affordance use amongst professional tweets. The results showed that chemists (0.27) had the highest mean of tweets with hashtags, while philosophers (0.09) had the lowest. As with personal tweets, sociologists had the highest mean of tweets with URLs, whereas professors in chemistry (0.11) and physics (0.11) had the lowest mean. The results examining user mentions showed that biology professors (0.76) had the highest mean of tweets with at least one user mention, while chemists (0.62) had the lowest. Professors in biology (0.26) had the highest mean of retweets and that physicists (0.20) had the lowest mean of tweets that were retweets. A $\chi^2$ test for association was conducted between department and affordance use. All expected cell frequencies were greater than five. There was a statistically significant association between department and hashtag use, $\chi^2(7\ n=6,810)=0.129$, p=0.0005, Cramer's V=0.129, between department and URL use, $\chi^2(7\ n=6,810)=0.177$, p=0.0005, Cramer's V=0.177, between department and user mention use, $\chi^2(7\ n=6,810)=0.166$, p=0.0005, Cramer's V=0.166, and between department and retweeting, $\chi^2(7\ n=6,810)=0.132$, p=0.0005, Cramer's V=0.132.

*4.3 Phase Three: Follow-up Survey and Tweet Categorization*
The 57 professors who responded to the follow-up survey were shown eight different affordances available in Twitter and asked to select ones they used in personal and professional tweets (as shown in Figure 4). URLs and mentions were chosen to frame professional tweets (85 and 85 percent, respectively) almost twice as often than for framing personal tweets (42 and 54 percent, respectively). Hashtags were chosen by 78 percent of respondents to frame professional tweets, whereas only 42 percent chose them for personal tweets; a similar trend was observed for retweets, where 80 percent chose them to frame professional tweets as compared to only 44 percent for personal tweets. The two affordances that were chosen to frame personal tweets at a higher percentage than those chosen to be used in professional tweets were media and emoticons; media was chosen 2 percent more for personal tweets than professional tweets (56 and 54 percent, respectively) and emoticons were chosen 31 percent of the time for personal tweets compared to only 14 percent in professional tweets.

Respondents were also asked if they had observed any of their own tweets being misinterpreted as personal when they were intended as professional or vice-versa and 15 percent (n=62) acknowledged they had had tweets that were misinterpreted in one way or the other, or as Goffman (1974) would argue, they had been misframed.

Results from the tweet categorization question are shown in Table III. A total of 255 tweets were presented to the respondents (n=51). Each respondent was shown five of their own random tweets; two of the tweets were personal and three were professional (as categorized by the turkers). The 51 scholars were made up of 15 assistant professors, 19 associate professors, and 17 full professors.

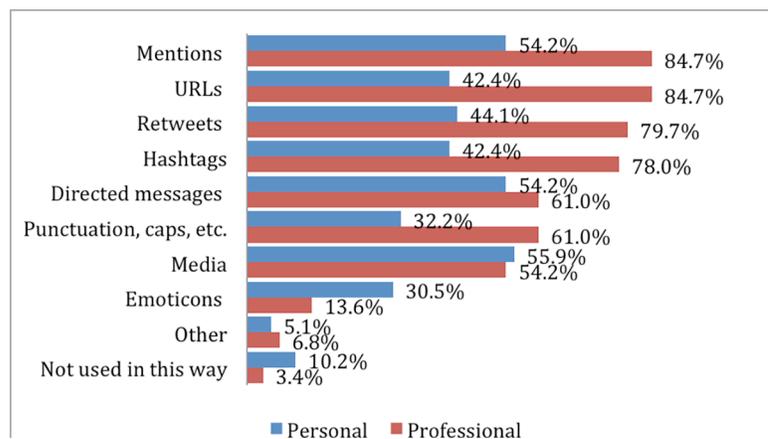

**Figure 4** Scholars selecting affordances to frame tweets as personal or professional

Cohen's κ was run to determine if there was agreement between the professor and the three turkers categorization of tweets as either personal or professional (see Table IV). There was fair agreement between the two judgements, κ=0.26. The problem lies with the perception of tweets intended to be professional (only 43 percent agreement as seen in Table III); when a turker categorized the tweet as personal (102 tweets total), respondents categorized the tweet as professional (58 tweets incorrectly categorized equaling 59 percent of all personal tweets).

Table 3 Results from scholar and turker coding agreement question

|  | Personal Tweets | Professional Tweets | Total |
|---|---|---|---|
| Turkers (3 out of 3) | 102 | 153 | 255 |
| Professors | 44 | 125 | 169 |
|  | 43% Agreement | 82% Agreement | 62% |

This problem of distinguishing between personal and professional tweets is at the heart of the difficulties faced by scholars, university administrators, and other organizations. When scholars' tweets are misinterpreted by the public or, to use Goffman's (1974) terms, when the audience frames a tweet in a specific way that was unintended by the tweeter, problems can occur (as discussed in the Introduction). For this work it is important to note that the small inter-rater reliability result from phase three in which scholars categorized their own tweets as personal or professional was expected and that this result should not diminish the overall interpretations of this study. What is important in these findings is that audience members are misinterpreting scholars' tweets and that affordance use in tweets may be helpful in framing the tweet in a specific way so that the audience correctly frames each tweet in the way the scholar intended.

Table 4 Cohen's kappa results for professors categorizing tweets showing the number of agreements with turkers.

|  | **OBSERVED AGREEMENT** | Turkers | |  | **EXPECTED AGREEMENT** | Turkers | |
|---|---|---|---|---|---|---|---|
|  |  | Personal | Professional |  |  | Personal | Professional |
| Professor | Personal | **44** | 28 | Professor | Personal | **29** | 43 |
| Professor | Professional | 58 | **125** | Professor | Professional | 73 | **110** |
|  |  |  |  |  |  |  | **Cohen's kappa = 0.26** |

## 5. Discussion and Conclusion

Social media metrics researchers attempt to identify and distinguish between the dissemination of, and engagement with, scientific content by measuring the traces (Cronin, 2001) left in social media environments by events related to scholarly communication (e.g. measuring a tweet containing a URL with a DOI to an academic paper). One significant problem with blindly measuring these events is that it is problematic for researchers to distinguish between personal and professional tweets, which makes it difficult to speak to whether a scholar is communicating about science in a personal or professional way and to decide whether or not this should be considered as academic output or impact. This work has taken a first step in attempting to identify differences in affordance use between personal and professional tweets. While there are other factors that are taken into consideration by audience members when reading a tweet including the tweet message itself, the time it was sent, the amount of responses it has received, and the familiarity of the audience members with the scholar, this work has identified framing strategies that should assist social media metric researchers to better differentiate between personal and professional communications in this context.

There are four main limitations that will be discussed here: study population, survey instrument, low survey response rate, and low Cohen's κ. The phase one survey invited scholars to talk about scholarly communication in Twitter. Therefore it potentially recruited one kind of scholar – those who have experience with Twitter and social media and who may be interested in scholarly communication. The second limitation involved the survey instrument; one disadvantage of the initial survey was the need to send invitations to scholars in two groups due to the limitations of the Qualtrics mail functionality. In addition, a shorter survey may have ensured greater survey response rate by decreasing the amount of time participants needed to answer all the questions. While e-mail reminders were sent to the population reminding them to fill out the phase one survey and the design of the survey was condensed to only 19 questions, this work was limited by a low survey response rate of 8.5 percent, which limits the generalizability of the results. Lastly, a fair rating was obtained using Cohen's κ for the inter-rater reliability measure in phase three of this work. While expected, some may consider this to limit the reliability of the data obtained from this phase of the work.

Affordance use by scholars in Twitter was shown to vary across personal and professional tweets based on department, gender, academic age, age, and Twitter activity. A difference in department affiliation suggests that there may be different social norms and framing behaviors by research area. This is intuitive as publication activity and general academic norms and behavior also vary by department (Piro, et al., 2013; Sugimoto & Cronin, 2012) and previous studies found disciplinary differences regarding Twitter uptake and activity to journal articles (Costas et al., 2014; Haustein, et al., 2014b). There were gender differences in affordance use across personal and professional tweets: women were found to have used affordances more than men in professional tweets and men had used affordances more in personal tweets.

The analysis also revealed that scholars use affordances differently at the beginning, middle, and late stages of their career. It was found that affordance use also varied when looking at academic age, age, and Twitter activity. As Gibson (1977) noted in his original conceptualization of an affordance, its use in a niche depends on an agent's ability to recognize the existence of the affordance. This could be one explanation of why affordance use varies as Twitter activity increases. Other differences could occur because social norms and rules inherent in offline communication have been brought to bear on communication activities within the Twitter environment and thus might influence how affordances in Twitter are used. To better frame these differences, it was useful to examine the framing patterns of these tweets as described by Goffman's theoretical model.

Framing behaviors were found on Twitter that distinguished personal and professional tweets. It was found that users tended to use certain affordances of Twitter more frequently than others when communicating in a personal or professional way. This evidence suggests that framing takes place in Twitter and that affordance use has a role in this behavior. The differences in affordance use between the personal and professional tweets indicate that there are patterns and distinctions to be made and a better understanding of why and how the scholars chose to use each affordance for specific types of tweets will help further identify patterns that can be used by researchers, universities, organizations, and the audience to better frame the communication.

To the author's knowledge, this work was one of the largest examinations to date of professors using Twitter (with Holmberg and Thelwall, 2014 reporting results of a study of 447 researchers). It is difficult to identify professors using Twitter and this method presented a unique way of identifying scholarly Twitter accounts. In addition, this is also one of the first studies to make use of AMT to categorize this amount of tweets in this manner. Future work should apply the suggested hybrid theoretical framework to a larger set of researchers and cover other academic departments to gather more substantial evidence to identify the framing behaviors used by scholar when they tweet personally and professionally in this environment.

**References**

Baldwin, R.G. (1998), "Technology's impact on faculty life and work", New Directions for Teaching and Learning, Vol. 1998 No. 76, pp. 7-21.

Bar-Ilan, J., Haustein, S., Peters, I., Priem, J., Shema, H. and Terliesner, J. (2012), "Beyond citations: Scholars' visibility on the social web", Proceedings of the 17th International Conference on Science and Technology Indicators, Montreal, available at: http://arxiv.org/abs/1205.5611 (accessed January 24, 2014).

Berrett, D. (2010), "ESU professor suspended for comments made on Facebook page", Pocono Record, February 26, available at: www.poconorecord.com/apps/pbcs.dll/article?AID=/20100226/NEWS/2260344 (accessed November 15, 2014).

Bowman, T.D., Demarest, B., Weingart, S.B., Simpson, G.L., Lariviere, V., Thelwall, M. and Sugimoto, C.R. (2013), "Mapping DH through heterogeneous communicative practices", Digital Humanities 2013, July 16-19.

boyd, D. (2011), "Social network sites as networked publics: affordances, dynamics, and implications", in Papacharissi, Z. (Ed.), A Networked Self: Identity, Community and Culture on Social Network Sites, Routledge, New York, NY, pp. 39-59. 368 boyd, d. and Ellison, N.B. (2008), "Social network sites: definition, history, and scholarship", Journal of Computer-Mediated Communication, Vol. 13 No. 1, pp. 210-230.

Costas, R., Zahedi, Z. and Wouters, P. (2014), "Do altmetrics correlate with citations? Extensive comparison of altmetric indicators with citations from a multidisciplinary perspective", Journal of the Association for Information Science and Technology, doi: 10.1002/asi.23309, early view.

Cronin, B. (2001), "Bibliometrics and beyond: some thoughts on web-based citation analysis", Journal of Information Science, Vol. 27 No. 1, p. 1-7.

Davis, C.H.F. III, Deil-Amen, R., Rios-Aguilar, C. and Canche, M.S. (2012), "Social media and higher education: a literature review and research directions", report Printed by the University of Arizona and Claremont Graduate University, available at: http://works.bepress.com/hfdavis/2/

Duque, A. and Pérez, M. del M. (2013), "Contribución de twitter a la mejora de la comunicación estratégica de las universidades latinoamericanas", RUSC, Vol. 10 No. 2, pp. 236-251.

Gibson, J.J. (1977), "The theory of affordances", in Shaw, R. and Bransford, J. (Eds), Perceiving, Acting, and Knowing: Toward an Ecological Psychology, Lawrence Erlbaum, Hillsdale, NJ, pp. 127-143.

Gilpin, D. (2011), "Working the twittersphere: microblogging as professional identity construction", in Papacharissi, Z. (Ed.), A Networked Self: Identity, Community and Culture on […], Routledge, pp. 232-250.

Goffman, E. (1959), The Presentation of Self in Everyday Life, 1st ed., Anchor Books, New York, NY.

Goffman, E. (1974), Frame Analysis: An Essay on the Organization of Experience, Northeastern University Press, Boston, MA, p. 586.

Hank, C., Tsou, A., Sugimoto, C.R. and Pomerantz, J. (2014), "Faculty and student interactions via Facebook: policies, preferences, and practices", It – Information Technology, Vol. 56 No. 5, pp. 216-223.

Haustein, S., Holmberg, K., Bowman, T.D. and Larivière, V. (2014a), "Automated arXiv feeds on Twitter: on the role of bots in scholarly communication", 19th Nordic Workshop on Bibliometrics and Research Policy.

Haustein, S., Peters, I., Bar-Ilan, J., Priem, J., Shema, H. and Terliesner, J. (2013), "Coverage and adoption of altmetrics sources in the bibliometric community", arXiv, Digital Libraries, pp. 1-12.

Haustein, S., Peters, I., Sugimoto, C.R., Thelwall, M. and Larivière, V. (2014b), "Tweeting biomedicine: an analysis of tweets and citations in the biomedical literature", Journal of the American Society for Information Science and Technology, Vol. 65 No. 4 pp. 656-669.

Haythornthwaite, C. and Wellman, B. (1998), "Work, friendship, and media use for information exchange in a networked organization", Journal of the American Society for Information Science, Vol. 49 No. 12, pp. 1101-1114.

Aslib Journal of Information Management 67(3)Hemphill, L., Culotta, A. and Heston, M. (2013), Framing in Social Media: How the US Congress Uses Twitter Hashtags to Frame Political Issues, August 28, pp. 30, available at SSRN: http://ssrn.com/abstract=2317335 or http://dx.doi.org/10.2139/ssrn.2317335

Herman, B. (2014), "Steven Salaita twitter scandal: university offers settlement, but free speech questions linger", International Business Times, September 5, available at: http://www.ibtimes.com/steven-salaita-twitter-scandal-university-offers-settlement-free-speech-questionslinger-1678854 (accessed November 15, 2014).

Hoffmann, C.P., Lutz, C. and Meckel, M. (2014), "Impact factor 2.0: applying social network analysis to scientific impact assessment", 2014 47th Hawaii International Conference on System Sciences, IEEE Press, Waikoloa, HI, pp. 1576-1585.

Holmberg, K. and Thelwall, M. (2014), "Disciplinary differences in twitter scholarly communication", Scientometrics, Vol. 101 No. 2, pp. 1027-1042.

Ingeno, L. (2013), "Outrage over professor's twitter post on obese students", Inside Higher Ed, June 4 (accessed November 15, 2014).

Jaschik, S. (2014), "Out of a job", Inside Higher Ed, August 6, available at: http://www.insidehighered.com/news/2014/08/06/u-illinois-apparently-revokes-job-offer-controversial-scholar (accessedNovember 15, 2014).

Lazer, D., Pentland, A., Adamic, L., Aral, S., Barabasi, A.L., Brewer, D., Christakis, N., Contractor, N., Fowler, J., Gutmann, M., Jebara, T., King, G., Macy, M., Roy, D. and Van Alstyne, M. (2009), "Life in the network: the coming age of computational social science", Science, Vol. 323 No. 5915, pp. 721-723.

Leary, M.R. and Kowalski, R.M. (1990), "Impression management: a literature review and two-component model", Psychological Bulletin, Vol. 107 No. 1, pp. 34-47.

Lough, T. and Samek, T. (2014), "The digital future of education international review of information ethics", International Review of Information Ethics, Vol. 21 No. 7, pp. 45-53.

Meyrowitz, J. (1990), "Redefining the situation: extending dramaturgy into a theory of social change and media effects", in Riggons, S. (Ed.), Beyond Goffman: Studies on Communication, Institution, and Social Interaction, Mouton de Gruyter, New York, NY, pp. 65-98.

Miller, H. (1995), "The presentation of self in electronic life: Goffman on the internet", Embodied Knowledge and Virtual Space Conference, London, p. 8.

Mitzlaff, F., Atzmueller, M., Stumme, G. and Hotho, A. (2013), "Complex networks IV", in Ghoshal, G., Poncela-Casasnovas, J. and Tolksdorf, R. (Eds), Complex Networks IV, Studies in Computational Intelligence, Springer Berlin Heidelberg, Berlin, Heidelberg, Vol. 476, pp. 13-25.

Murthy, D. (2013), Twitter: Social Communication in the Twitter Age, Polity Press, Cambridge, p. 220.

Papacharissi, Z. (2011), "Conclusion: a networked self", in Papacharissi, Z. (Ed.), A Networked Self: Identity, Community and Culture on Social Network Sites, Routledge, New York, NY, pp. 304-318.

Piro, F., Aksnes, D. and Rørstad, K. (2013), "A macro analysis of productivity differences across fields: challenges in the measurement of scientific publishing", Journal of the America…, Vol. 64 No. 2, pp. 307-320.

Priem, J. (2010), "I like the term #articlelevelmetrics, but it fails to imply *diversity* of measures. Lately, I'm liking #altmetrics", Twitter, September 28, available at: https://twitter.com/jasonpriem/status/25844968813 (accessed December 21, 2014).

Priem, J. (2014), "Altmetrics", in Cronin, B. and Sugimoto, C.R. (Eds), Beyond Bibliometrics: Harnessing Multidimensional Indicators of Scholarly Impact, MIT Press, Cambridge, MA, pp. 263-288.

Priem, J., Groth, P. and Taraborelli, D. (2012), "The altmetrics collection", in Ouzounis, C.A. (Ed.), PloS One, Public Library of Science, Vol. 7 No. 11, p. e48753.

Priem, J., Taraborelli, D., Groth, P. and Neylon, C. (2010), "Altmetrics: a manifesto", available at: http://altmetrics.org/manifesto/ (accessed December 21, 2014).

Rothschild, S. and Unglesbee, B. (2013), "Kansas university professor receiving death threats over NRA tweet", The Dispatch, September 24, available at: www.shawneedispatch.com/news/2013/sep/24/kansas-university-professor-receiving-death-threat/ (accessed November 14, 2014).

Schroeder, R., Power, L. and Meyer, E.T. (2011), "Putting scientometrics 2.0 in its place [vO]", Altmetrics11. Tracking Scholarly Impact on the Social Web. An ACM Web Science Conference 2011 Workshop, Koblenz, June 14-15, available at: http://altmetrics.org/workshop2011/schroeder-v0/

Sugimoto, C., Hank, C., Bowman, T. and Pomerantz, J. (2015), "Friend or faculty: social networking sites, dual relationships, and context collapse in higher education", First Monday, Vol. 20 No. 3, doi:10.5210/fm.v20i3.5387.

Sugimoto, C.R. and Cronin, B. (2012), "Bibliometric profiling: an examination of multifaceted approaches to scholarship", Journal of the American Society for Information Science and Technology, Vol. 63 No. 3, pp. 450-468.

Veletsianos, G. (2012), "Higher education scholars' participation and practices on twitter", Journal of Computer Assisted Learning, Vol. 28 No. 4, pp. 336-349.